\documentclass[prl,aps,twocolumn,showpacs]{revtex4}
\usepackage{epsfig}
\usepackage{amssymb}
\usepackage{amsmath}
\usepackage{graphicx}
\usepackage{amsfonts}
\usepackage{color}

\begin{document}

\title{Magnetic and Superconducting Properties of   FeAs-based High-Tc Superconductors with Gd}

\author{E.~P.~Khlybov$^{a,f}$, O.~E.~Omelyanovsky$^{b,c}$,
A.~Zaleski$^d$, A.~V.~Sadakov$^{b,c,f}$, D.~R.~Gizatulin$^{b,e}$,
L.~F.~Kulikova$^a$, I.~E.~Kostyleva$^{a,f}$, V.~M.~Pudalov$^{b,c}$
}

\affiliation{$^{a}$ Institute for High Pressure Physics, RAS,
Troitsk, Moscow region, 142190 Russia}
\affiliation{$^{b}$P.~N.~Lebedev Physical Institute, RAS,Moscow,
119991 Russia} \affiliation{$^{c}$ P.~N.~Lebedev Research Centerin
Physics, Moscow, 119991 Russia} \affiliation{$^{d}$ Institute of
Low Temperatures and  Structure Research PAN, Wroclaw, 50-950
Poland} \affiliation{$^{e}$Moscow Institute for Physics and
Technology, Russia} \affiliation{$^{f}$ International Laboratory
of High magnetic fields and Low Temperatures, Wroclaw, 53-421
Poland }

\begin{abstract}We report on successful synthesis under high
pressure of a series of polycrystalline GdFeAs O$_{1-x}$F$_x$
high-T$_c$ superconductors with different   oxygen deficiency
$x=0.12 \div 0.16$ and also with no fluorine. We have found that
the high-pressure synthesis technique is crucial for obtaining
almost single-phase superconducting materials:  by synthesizing
the same compounds with no pressure in ampoules we obtained
non-superconducting materials with an admixture of incidental
phases. Critical temperature for all the materials was in the
range 40 to 53\,K. The temperature derivative of the critical
field  $dH_{c2}/dT$ is remarkably high, indicating potentially
high value of the second  critical field  H$_{c2} \sim 130$\,T.
\end{abstract}


\maketitle Recent discovery of a new class of high temperature
superconductors based on iron \cite{kamihara08,takahashi_nature08}
came as a big surprise to the theory; it  has sparked vast
interest and stimulated intensive experimental and theoretical
research of the superconductivity in this class of material.  The
first discovered high temperature pnictide superconductor
LaFeAsO$_{1-x}$F$_x$ had a critical temperature $T_c = 26$\,K
\cite{kamihara08},   the substitution of La by Se, Sm, Pr, Nd was
shown to increase  critical temperature \cite{chen_nature08}. The
highest  $T_c\approx54$\,K was  found for SmFeAsO$_{1-x}$ with
optimized  oxygen deficiency $x=0.2$ \cite{liu08}.

High interest to these compounds is stimulated by the theoretical
suggestion that the  superconductivity in  iron-based
superconductors is unconventional and mediated by spin
fluctuations \cite{mazin08,sadovskii_UFN,chen_PRL08}.  Another
point of interest is an unusual combination of magnetic ordering
and superconducting pairing in the same material. The existence of
the spin density wave ordering in the  superconducting phase gives
rise to the unusual $s_{\pm}$ type pairing symmetry
\cite{sadovskii_UFN,golubov_08,parker_09,chen_09}.   The structure
of the so called 1111-compounds RZFeAsO$_{1-x}$F$_x$ (RZ=La, Ce,
Pr, Nd, Sm, Gd, Tb, Dy) consists of alternating RZO and FeAs
layers stacking along the $c$-axis \cite{izyumov_UFN}. Doping of
the compounds with $n-$ or $p-$ type carriers is achieved by
partial substitution of O or RZ \cite{izyumov_UFN,ivanovskii_UFN}.

Due to its high critical temperature ($\sim54$\,K), Sm-based
FeAs-superconductor  is studied in most detail. In the current
paper we focus on much less studied Gd-based FeAs material. There
were reports about $T_c=36$\,K for  GdFeAsO$_{0.83}$F$_{0.17}$ \cite{cheng_08} and
$T_c=54$\,K for  GdFeAsO$_{0.7}$ \cite{miyazawa09}. In this
paper we describe method of synthesis of almost single-phase
superconducting polycrystals of this material with various amount
of oxygen and fluorine.  For optimally doped GdFeAs-based
superconductor, we obtained  $T_c =53$\,K and estimated  $H_{c2}
\sim 130$\,T.

{\bf Synthesis.} Two methods for solid state synthesis of
polycrystalline pnictides at high temperatures were  mentioned in
literature \cite{izyumov_UFN,ivanovskii_UFN,karpinski_09}:
synthesis  in evacuated quartz ampoules, and synthesis at high
pressure.  We tested both methods and found that the reproducible
high quality, almost single-phase material is obtained only in
high pressure synthesis. The starting materials for synthesis were
high purity  chips  of Gd and As (99.9\%) and powders  FeF$_3$,
Fe, and Fe$_2$O$_3$ (99.99\%). Initially, the chips of Gd and As
were placed in an evacuated quartz ampoule and heated at
$T=1050$$^\circ$C during 24h. Purity of the synthesized GdAs
phase was tested by the powder  X-ray diffraction. The resulting
GdAs powder,  FeF$_3$, Fe, and Fe$_2$O$_3$ powders were mixed
together with the nominal stoichiometric ratio, and  then pressed
into  pellets (3\,mm od and 3mm height).\\
{\bf  For high pressure synthesis} we used  ``Conac-28'' high
pressure apparatus \cite{conac}. The pellets were inserted in
boron nitride crucible and synthesized  at pressure of 50\,kb  and
temperature 1350$^\circ$C during 60\,min. Further, the temperature
was either (i) decreased to 1200$^\circ$C in 60 min and then
heating was switched off, or switched off right after 60min-stage
of synthesis. The X-ray diffraction (XRD) pattern (Fig.~1)
demonstrates  that the resulting substance is practically a pure
single-phase polycrystalline material. \\
{\bf For  ampoule synthesis} the pellets (prepared as described
above)  were inserted in fused quarts  evacuated ampoules and were
sintered in oven at temperature 1180$^\circ$C during 24 hours.
Samples produced by the ampoule synthesis showed similar XRD
pattern (see Fig.~1), however  additional peaks evidence for a
noticeable admixture of incidental phases. All samples synthesized
in ampoules were non-superconducting. Re-grinding the sintered
pellets and repetition of the synthesis   in ampoules did not help
to produce the desired superconducting phase. However, after the
pellets were re-synthesized under high pressure, as described
above, the materials became superconducting with XRD pattern and
other properties  similar to those for the materials synthesized
at high pressure directly from powders. We conclude that  the
ampoule  reaction technique is not suitable for obtaining
desirable single-phase GdFeAsOF-compound.

X-ray diffraction was taken at room temperature using
Mo-$K_\alpha$ radiation. Almost all characteristic peaks in the
spectra (Fig.~1) are identified and evidence for a nearly
single-phase polycrystalline material.   The XRD pattern in Fig.~1
is presented for a typical sample   synthesized by the HP
technique and, for comparison, for another sample of a nominally
similar composition but synthesized by the  ampoule technique.

\begin{figure}[ht]
\begin{center}
\includegraphics[width=.54\textwidth]{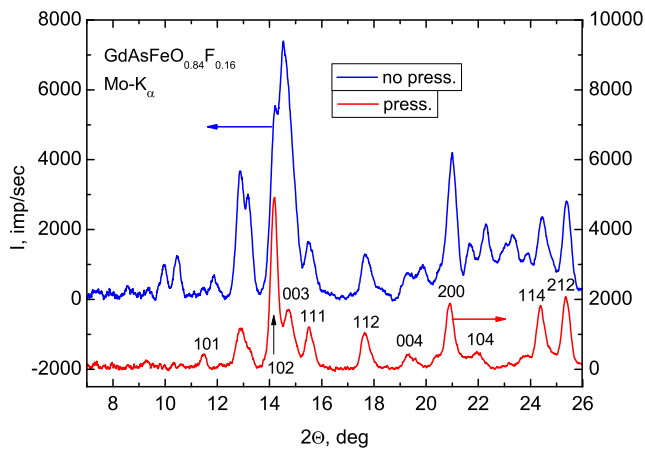}
\caption{Fig. 1. X-ray diffraction pattern for two samples of the nominally same composition and fabricated by ampoule (A) and high pressure (HP) synthesis. Monotonic background is subtracted in both cases. Peaks are indexed according to $P4/nmm$ symmetry of the lattice \protect\cite{ivanovskii_UFN}.}
\end{center}
\label{fig:F1}
\end{figure}
\vspace{0.1in}
For the
samples synthesized at high pressure, we
performed measurements of the magnetic susceptibility (by
ac-technique, at $\approx 900$\,Hz, and with modulation amplitude
0.1\,Oe) and resistivity (by standard four-probe technique).
Figure 2 shows temperature dependence of the susceptibility in
zero magnetic field, measured for four samples with various
content of oxygen  and fluorine.  All samples exhibit a sharp
superconducting transition with a critical temperature $T_c$
ranging from  35 to 50\,K. The  critical temperature which is
higher than that for some high temperature superconductors, such
as MgB$_2$ ($T_c=39$K) and for La$_{2-x}$Ba$_{x}$CuO$_4$
($T_c=36$K), but lower than for YBa$_2$Cu$_3$O$_{7-x}$
($T_c=90$K), is the evidence that the material belongs to the
class of high temperature superconductors (HTSC).

For comparison, the insert to Fig.~2 shows ac-susceptibility
measured for three samples  synthesized in ampoules. Symbols on
the main panel and on the insert refer to nominally the same
amount of oxygen and fluorine. Clearly, the samples produced by
ampoule synthesis are not superconducting, at least above 2\,K,
and show an antiferromagnetic type dependence with Neel
temperature $\Theta_N \approx 17$\,K. It is worthy of note that the
samples synthesized at high pressure in the normal state exhibit a
paramagnetic temperature dependence (though almost
indistinguishable from antiferromagnetic type dependence in the
available temperature range).

\begin{figure}[ht]
\begin{center}
\includegraphics[width=.6\textwidth]{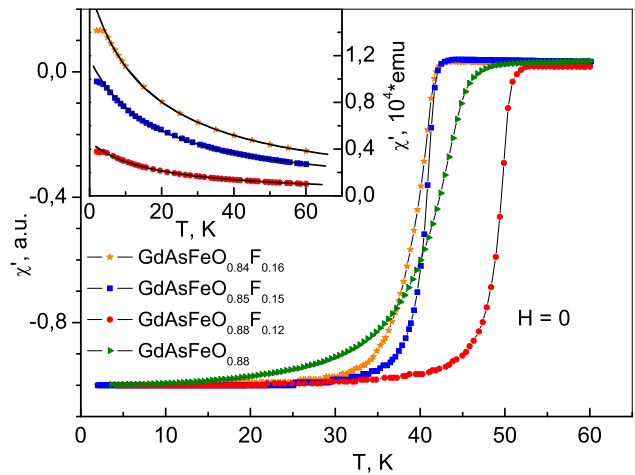}
\caption{Fig. 2. Temperature dependence of the ac-susceptibility measured at  zero magnetic field for four samples. The isert shows susceptibility for three samples with nominally the same content, but synthesized in ampoules.}
\end{center}
\label{fig:F2}
\end{figure}

As seen from Fig.~2, the  GdFeAsO$_{0.88}$F$_{0.12}$  sample has the highest $T_c\approx 50$\,K among the four synthesized  samples. For this ``optimally''  (or almost optimally) doped  sample we present in Fig.~3 temperature dependence of the resistance, measured in various magnetic fields. As magnetic field  increases, the critical temperature decreases and the width of the transition increases; as a result, the beginning of the superconducting transition stays almost unchanged. Such behavior is typical for type II superconductors, and particularly, for SmFeAsO$_{0.7}$F$_{0.29}$ \cite{karpinski_09}.  The critical temperature determined from the temperature dependence of resistance  is 52.5\,K, noticeably higher than that determined from the $\chi^\prime(T)$ measurements. Such difference is characteristic for type II superconductors.

\begin{figure}[ht]
\begin{center}
\includegraphics[width=.52\textwidth]{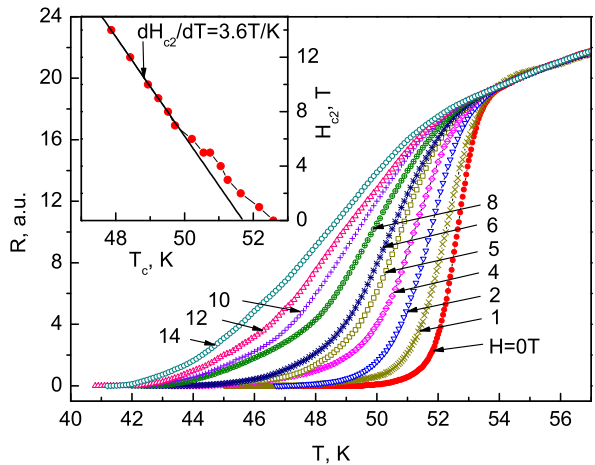}
\caption{Fig.~3. Temperature dependence of the resistance for various magnetic fields. Labels next to curves designate external magnetic field value in Tesla. The insert  shows temperature dependence of the critical field $H_{c2}$, determined from data on the main panel at the middle of the transition. }
\end{center}
\label{fig:F3}
\end{figure}

In order to evaluate relative content of the superconducting phase in the sample, we compare in Fig.~4 the temperature dependence of the magnetic susceptibility measured for different conditions:\\
(a) ``Zero field cooling'' (ZFC) - the sample was cooled in zero magnetic field, further, the required magnetic field was applied at the lowest temperature (2K)  and  the ac-susceptibility was measured  during warming the sample in the given field.\\
(b) ``Field cooling'' (FC) -  the required magnetic field was applied at a temperature  much higher   than $T_c$  and the ac-susceptibility was measured during cooling the sample in a given magnetic field.

\begin{figure}[hc]
\begin{center}
\includegraphics[width=.55\textwidth]{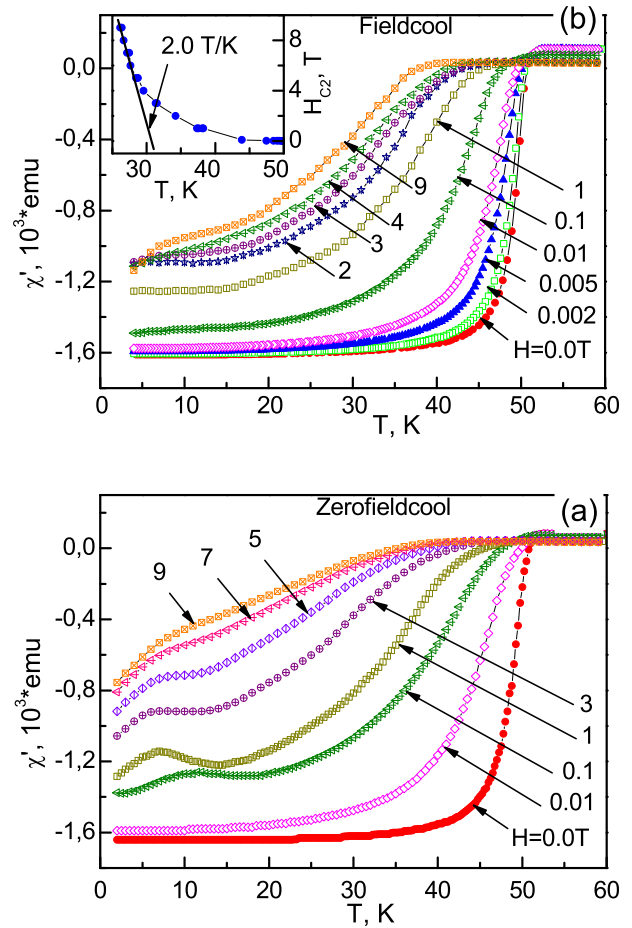}
\caption{Fig.~4. Temperature dependence of the real part of the magnetic susceptibility $\chi^\prime(T)$ measured (a) for ``zero field cooling'' (ZFC) and (b) for ``field cooling'' (FC) in various magnetic fields. Field values are indicated on the figure. The insert  shows temperature dependence of the critical field $H_{c2}$, determined from data on the main panel at the middle of the transition.}
\end{center}
\label{fig:F4}
\end{figure}

Comparing Figs. 4a and 4b, one can see that in low magnetic fields
($<0.01$\,T), the  curves ZFC-$\chi^\prime(T)$ and
FC-$\chi^\prime(T)$ are almost the same. This is in a sharp
contrast to previous  data for polycrystalline GdFeAs based
superconductors from Refs. \cite{cheng_08,miyazawa09}, where
almost 1:20 ratio of the ZFC to FC data was reported.  Even for
nominally single-crystals of SmFeAsO$_{0.6}$F$_{0.35}$
\cite{karpinski_09}, the ratio is an order of 1:20. We believe,
that the coincidence of ZFC and FC data (Figs. 4a and 4b)
indicates that the superconducting phase in our sample  has a bulk
character and constitutes almost 100\% of the sample volume. In
higher fields, the ZFC-$\chi^\prime(T)$ and FC-$\chi^\prime(T)$
behavior becomes different; we have currently no explanation for
the observed unusual difference. Interestingly, $\chi^\prime(T)$
measured in ZFC-regime demonstrates a wide peak at low
temperatures (below 20K) and in fields 0.1--9\,T. This  behavior
might be related with magnetic ordering of Gd ions. This effect
requires further studies.

Besides $T_c$, another important parameter is the second critical
field $H_{c2}$. The  $H_{c2}$ value is known to be high for
FeAs-type superconductors. As an example, for
SmFeAsO$_{0.7}$F$_{0.25}$ the measured derivative is
$dH_{c2}/dT=-(2\div4)$T/K  \cite{karpinski_09}; this leads to a
rough estimate $H_{c2}\sim 70\div140$T. For our ``optimally
doped'' GdFeAsO$_{0.88}$F$_{0.12}$ polycrystalline  sample, the
temperature derivative of the critical field $dH_{c2}/dT$ was
found from the measured $R(T,H)$  and  $\chi^\prime(T,H)$
temperature dependencies. In the inserts to Fig.~3 and Fig.~4 we
plotted the  $H_{c2}(T)$ dependence, determined at the middle of
the superconducting transition from both, $R(T)$ and
$\chi^\prime(T)$ curves. From these data   we obtain  the slope
$|dH_{c2}/dT|$  equal to 2 and 3.6\,T/K, respectively. To compare
with other reported data for oxypnictide superconductors, we use
the latter estimate.   The slope is a factor of four larger  than
for LaFeAsO$_{1-x}$F$_x$ \cite{La_Hc2}, and comparable to that for
optimally doped SmFeAs-based superconductors \cite{karpinski_09}.
Using the conventional  Werthamer-Helfand-Hohenberg BCS theory for
type II superconductors \cite{WHH}  we obtain an estimate $H_{c2}
=-0.693T_c \left(\partial H_{c2}/\partial T\right)|_{T_c} \approx
130$T, comparable to $H_{c2}$ for YBCO (though with
$T_{c}=92$\,K). This estimate, of  course, will be  affected by
the paramagnetic limit which we currently don't now. Nevertheless,
potentially high values of the critical field, up to 130\,T make
the Gd-FeAs superconductor interesting for practical high-field
applications.

{\em In summary}, we elaborated high pressure synthesis method and
synthesized several GdFeAsO$_{1-x}$F$_{x}$ polycrystalline samples
with  different content of oxygen and
fluorine; the synthesized materials have critical temperature  in
the range $38-52$\,K. We also found that the high-temperature
synthesis in ampoules  is not suitable for obtaining single-phase
superconductors of this composition.  The highest $T_c\approx53$K
was found for GdFeAsO$_{0.88}$F$_{0.12}$. The  evidence for almost
single-phase content of the superconductors is provided by the X-ray diffraction
and by comparison of the temperature dependences of the magnetic
susceptibility measured in  ``zero field cooling'' and ``field
cooling'' conditions. The ``optimally doped'' samples with
$x=0.12$ demonstrate high value of the derivative $dH_{c2}/dT$,
which leads to an estimate for $H_{c2}(0)$ up to $\sim 130$\,T.

The authors are thankful to E.~V.~Antipov, S.~M.~Kazakov, and
E.~G.~Maksimov  for discussions, and V.~L.~Ginzburg for his
interest to the present work. The work was partially supported by
grants from RFBR (09-02-12206, 09-02-01370), by Federal agency on
science and innovation (02.513.11.3378), and by Federal agency on
education.

\end{document}